\documentstyle[prl,aps,epsfig,amssymb,multicol]{revtex}
\begin{document}

\draft
\preprint{SNUTP 98-044}
\title{Cotunneling Transport and Quantum Phase Transitions in
Coupled Josephson-Junction Chains with Charge Frustration}

\author{Mahn-Soo Choi$^1$, M.Y. Choi$^2$, Taeseung Choi$^1$ and
  Sung-Ik Lee$^1$}
\address{$^1$ Department of Physics, Pohang University of Science and
  Technology, Pohang 790-784, Korea}
\address{$^2$ Department of Physics and Center for Theoretical
  Physics, Seoul National University, Seoul 151-742, Korea}
\date{To appear in Phys.\ Rev.\ Lett.; cond-mat/9806365}
\maketitle

\begin{abstract}
We investigate the quantum phase transitions in two capacitively coupled
chains of ultra-small Josephson-junctions, where the particle-hole symmetry
is broken by the gate voltage applied to each superconducting
island.  Near the maximal-frustration line, cotunneling of the particles
along the two chains is shown to play a major role in the transport and to
drive a quantum phase transition out of the charge-density wave insulator,
as the Josephson-coupling energy is increased.  We also argue briefly that
slightly off the symmetry line, the universality class of the transition
remains the same as that right on the line, being driven by the
particle-hole pairs.
\end{abstract}

\pacs{PACS numbers: 74.50.+r, 67.40.Db, 73.23.Hk}

\newcommand\BfC { {\mathbb{C}} }
\newcommand\half { {\frac{1}{2}} }
\newcommand\hatC { \widehat{C} }
\newcommand\varN { {\cal N} }
\newcommand\varS { {\cal S} }
\newcommand\varE { {\cal E} }
\newcommand\ketl { \left| }
\newcommand\ketr { \right\rangle }
\newcommand\avgl { \left\langle }
\newcommand\avgr { \right\rangle }
\newcommand\tildeS { \widetilde{S} }
\newcommand\tildeH { \widetilde{H} }
\newcommand\indicator {\noindent{\bf\P}\quad}
\newcommand{\onecolm}{
  \end{multicols}
  \noindent\rule{0.5\textwidth}{0.1ex}\rule{0.1ex}{2ex}\hfill
}
\newcommand{\twocolm}{
  \hfill\raisebox{-1.9ex}{\rule{0.1ex}{2ex}}\rule{0.5\textwidth}{0.1ex}
  \begin{multicols}{2}
}

\begin{multicols}{2}


Systems of ultra-small tunnel junctions composed of metallic or
superconducting electrodes have been the source of a great number of
experimental and theoretical works~\cite{Graber92}.
Single charge (electron or Cooper pair) tunneling observed in
those systems demonstrates the remarkable effects of Coulomb blockade.
Especially, in Josephson-junction arrays, the charging energy in
competition with the Josephson-coupling energy further brings about the noble
effects of quantum fluctuations, which induce quantum phase transitions
at zero temperature~\cite{Cerdei96}.
Very recently, another fascinating manifestation of Coulomb blockade
has been revealed in capacitively coupled one-dimensional (1D) arrays of
metallic tunnel junctions \cite{Averin91a,Matter97}:
In such coupled chains, the major transport along both chains 
occurs via {\em cotunneling} of
the electron-hole pairs, which is a quantum mechanical process through an
intermediate virtual state.  Such a cotunneling transport leads to the
interesting phenomenon of the current mirror.

In capacitively coupled Josephson-junction chains, the
counterpart of the electron-hole pair is the particle-hole pair, i.e., the
pair of an excess and a deficit in Cooper pairs across the two chains.
Such particle-holes pairs, combined with the quantum fluctuations, have been
proposed to drive the insulator-to-superconductor transition
\cite{mschoi97p5}. 
Here it should be noticed that the particle-hole pair is stable only near
the particle-hole symmetry line; far away from the symmetry line,
it does not make the lowest charging-energy configuration anymore.
Moreover, in a single chain of Josephson junctions, 
breaking the particle-hole symmetry (by applying a gate voltage) is
known to change immediately the
universality class of the transition~\cite{Bruder93,Otterl94}.
Therefore, it is necessary to find another relevant cotunneling
process, if any, off the particle-hole symmetry line 
and to examine how the transitions change 
in coupled Josephson-junction chains.

As an attempt toward that goal,
we investigate in this paper the quantum phase transitions in two chains
of ultra-small Josephson-junctions, coupled capacitively with each other.
The particle-hole symmetry is
broken by the gate voltage applied to each superconducting island;
the resulting induced charge introduces {\em frustration} to the system.
Near the maximal-frustration line, cotunneling of the
particles along the two chains is found to play a major role 
in the transport and to
drive a quantum phase transition out of the charge-density wave (CDW)
insulator, as the Josephson-coupling energy is increased.
We also argue that slightly off the symmetry line, the universality class
of the transition remains the same as that right on the line, i.e., a
Berezinskii-Kosterlitz-Thouless (BKT) transition~\cite{Berezi71}, driven by
the particle-hole pairs.


We consider two 1D arrays, i.e., chains of Josephson junctions, 
each of which is characterized
by the Josephson coupling energy $E_J$ and the charging
energies $E_0\!\equiv\!e^2/2C_0$ and
$E_1\!\equiv\!e^2/2C_1$, associated with the
self-capacitance $C_0$ and the junction capacitance $C_1$,
respectively (see Fig.~\ref{ccjjcg:fig1}).
The two chains are coupled with each other via the
capacitance $C_I$, with which the electrostatic energy
$E_I\!\equiv\!e^2/2C_I$ is associated, while no
Cooper-pair tunneling is allowed between the two chains~\cite{ccjjcg:e1}.  
The intra-chain capacitances are assumed to be so small
($E_J\ll{}E_0,E_1$) that, without the coupling, each chain
would be in the insulating phase~\cite{Bradle84}.
We are interested in the limit where the coupling capacitance is
sufficiently large compared with the intra-chain capacitances,
$C_I\gg{}C_0,C_1$, i.e., $E_I\ll{}E_0,E_1$ [see Eq.~(\ref{ccjjcg:H_C1})
below].
On each superconducting island, external gate voltage $V_g$ is applied,
and accordingly, the external charge $n_g\!\equiv\!C_0 V_g/2e$ is induced,
with $e$ being the electric charge.
The external charge $n_g$ breaks the particle-hole symmetry of the system,
introducing charge frustration.
We restrict our discussion to two regions: near the particle-hole
symmetry line ($|n_g{-}\varN|\ll1/4$ with $\varN$ integer) and near the
maximal-frustration line ($|n_g{-}\varN{-}1/2|\ll1/4$), where
the properties of the system are severely different.
We further note the invariance with respect to the substitution
$n_g\to{}n_g+1$, and take $\varN=0$ without loss of generality.

The Hamiltonian describing the system is given by
\onecolm
\begin{equation}
H = 2e^2\sum_{\ell,\ell';x,x'}
  [n_\ell(x)-n_g]\BfC_{\ell\ell'}^{-1}(x,x')[n_{\ell'}(x')-n_g]
  - E_J\sum_{\ell,x}\cos[\phi_\ell(x)-\phi_\ell(x+1)]
  ,
\label{ccjjcg:H_QPM}
\end{equation}
\twocolm\noindent
where the number $n_\ell(x)$ of the Cooper pairs and
the phase $\phi_\ell(x)$ of the superconducting order parameter at site
$x$ on the $\ell$th chain ($\ell=1,2$)
are quantum-mechanically conjugate variables:
$[n_\ell(x),\phi_\ell(x')]=i\delta_{\ell\ell'}\delta_{xx'}$.
The capacitance matrix $\BfC$ in Eq.~(\ref{ccjjcg:H_QPM}) can be written in
the block form:
\begin{equation}
\BfC_{\ell\ell'}(x,x')
\equiv C(x,x')\left[ \begin{array}{cc}
    1\, & \,0 \\ 0\, & \,1
  \end{array} \right]
  + \delta_{x,x'}C_I\left[ \begin{array}{rr}
      1 & -1 \\ -1 & 1
    \end{array} \right]
\label{ccjjcg:CC}
\end{equation}
with the intra-chain capacitance matrix
\[
C(x,x')
\equiv C_0\delta_{xx'}+C_1\left[
    2\delta_{xx'}-\delta_{x,x'+1}-\delta_{x,x'-1}
  \right]
  .
\]
For simplicity, we keep only 
the on-site and the nearest neighbor interactions between the charges 
(i.e., $C_1/C_0\lesssim1$) 
although this is not essential in the subsequent discussion 
(as long as the interaction range is finite).  
With the block form of the
capacitance matrix in Eq.~(\ref{ccjjcg:CC}), the Hamiltonian can be
conveniently expressed as the sum
\begin{equation}
H=H_C^0+H_C^1+H_J
\end{equation}
with the components
\begin{eqnarray}
H_C^0
& \equiv & U_0\sum_x[n_+(x)-2n_g]^2
  + V_0\sum_x[n_-(x)]^2
\nonumber \\
H_C^1
& \equiv & U_1\sum_x[n_+(x)-2n_g][n_+(x{+}1)-2n_g]
  \nonumber\\&&\mbox{}
  + V_1\sum_xn_-(x)n_-(x{+}1)
  \label{ccjjcg:H_C1}\\
H_J
& \equiv & -E_J\sum_{\ell,x}\cos\left[\phi_\ell(x)-\phi_\ell(x+1)\right]
  \nonumber
  ,
\end{eqnarray}
where $n_\pm(x)\equiv{}n_1(x)\pm{}n_2(x)$ and the coupling strengths are
given by $U_0\simeq2E_0$, $U_1\simeq4(C_1/C_0)E_0$, $V_0\simeq{}E_I$, and
$V_1\simeq(C_1/C_I)E_I$.

The on-site charging energy term of the Hamiltonian $H_C^0$ in
Eq.~(\ref{ccjjcg:H_C1}) reveals clearly the crucial difference between the
charge configurations in the system near the maximal-frustration
line $n_g =1/2$ and those near the particle-hole symmetry line
$n_g =0$.
In the former region ($|n_g {-}1/2|\ll1$), 
the charge configurations which do not satisfy the
condition $n_+(x)=1$ (for all $x$) have a huge excitation gap
of the order of $E_0$.  
(Note that we are interested in the parameter regime $E_I,E_J\ll{}E_0,E_1$.) 
Furthermore, the ground states of $H_C^0$, separated
from the excited states by the gap of the order of $E_I$, have two-fold
degeneracy for each $x$, corresponding to $n_-(x)=\pm1$.
This degeneracy is lifted 
as the Josephson-coupling energy $E_J$ is turned on.  As a result,
it is convenient in this case to work within the reduced Hilbert space
$\varE_d$, where $n_+(x)=0$ and $n_-(x)=\pm1$ for each $x$.
In the latter region ($|n_g|\ll1/4$), on the other hand,
the low-energy charge configuration
should satisfy the condition $n_+(x)=0$ for all $x$.  Unlike the
former case, the ground state of $H_C^0$ is non-degenerate and forms a Mott
insulator characterized by $n_1(x)=n_2(x)=0$ for all $x$.  
As $E_J$ is turned on, the
ground state of $H_C^0$ is mixed with the states with $n_-(x)=\pm2$.
Accordingly, the relevant reduced Hilbert space is given by
$\varE_s$, where $n_+(x)=0$ and $n_-(x)=0,\pm2$ for all
$x$ (see also Ref.~\onlinecite{mschoi97p5}).


We first consider the region near the maximal-frustration lines, where we
project the Hamiltonian Eq.~(\ref{ccjjcg:H_QPM}) onto $\varE_d$, and
analyze the properties of the system near the
maximal-frustration line ($|n_g {-}1/2|\ll1/4$), based on the resulting
effective Hamiltonian.  Given the projection operator $P$ onto $\varE_d$,
the effective Hamiltonian up to the second order in $E_J/E_0$,
\begin{equation}
H_{\it eff}
\equiv P\left[
    H_{\it QPM}
    + H_J\frac{1-P}{E-H_C^0}H_J
  \right]P
  ,
\label{ccjjcg:proj}
\end{equation}
can be obtained via the standard procedure~\cite{Bruder93,Glazma97}.
Implementing explicitly the projection procedure, we get 
the effective Hamiltonian describing
a {\em single} spin-1/2 antiferromagnetic Heisenberg chain~\cite{ccjjcg:e2}
\begin{eqnarray}
H_{\it eff}
& = & \gamma J\sum_xS^z(x)S^z(x{+}1)
  \label{ccjjcg:AFHM} \\&&\mbox{}
  - \half J\sum_x\left[ S^+(x)S^-(x{+}1) + S^-(x)S^+(x{+}1) \right]
  , \nonumber
\end{eqnarray}
where the exchange interaction and the uniaxial anisotropy factor are given
by $J\equiv{}E_J^2/4E_0$ and $\gamma\equiv16\lambda^2E_I^2/E_J^2$,
respectively.  The pseudo-spin operators have been defined according to:
\begin{eqnarray}
&&S^z(x)
\equiv P\,\frac{n_1(x){-}n_2(x)}{2}\,P 
\nonumber \\
&&S^+(x)
\equiv Pe^{-i\phi_1(x)}(1-P)e^{+i\phi_2(x)}P \label{ccjjcg:S+} \\
&&S^-(x)
\equiv Pe^{-i\phi_2(x)}(1-P)e^{+i\phi_1(x)}P.
\nonumber
\end{eqnarray}
(Note the difference from the standard definition.)

The effective Hamiltonian in Eq.~(\ref{ccjjcg:AFHM}) includes 
contributions from several complex processes, back in the charge picture:
The first term in Eq.~(\ref{ccjjcg:AFHM}), which comes
from the projection $PH_C^1P$, simply describes the nearest-neighbor
interaction of the charges in the form $n_-(x)$.
On the other hand, the second term in Eq.~(\ref{ccjjcg:AFHM}), arising
from the second order expansion $PH_J\frac{1-P}{E-H_C^0}H_JP$, describes
the {\em cotunneling} process of two particles on different chains
by way of a quantum mechanical virtual state with energy of the order of
$E_0$.  Figure~\ref{ccjjcg:fig2} shows schematically a
particle at $x$ on one chain and another particle at $x{+}1$ (or $x{-}1$) 
on the other chain hopping at the same time in the opposite direction.  This
cotunneling process plays the major role in the charge transport along the
two chains, and drives the quantum phase transition in the system, 
as discussed below.

The antiferromagnetic Heisenberg chain described by 
Eq.~(\ref{ccjjcg:AFHM}) has been extensively studied~\cite{Auerba94},
and is known, from the Bethe ansatz solution 
or the Sine-Gordon theory~\cite{Fradki91},
to exhibit a quantum phase transition at $\gamma=1$: 
For $\gamma>1$, it belongs to the universality
class of the Ising chain in the renormalization group (RG) sense, 
and its ground state displays genuine
long-range order in the staggered magnetization,
i.e., $\avgl{(-1)^xS^z(x)S^z(0)}\avgr$ approaches a nonzero
constant as $x\to\infty$.  
This long-range order in the
staggered magnetization corresponds to the charge-density wave (CDW)
in the charge picture of the original problem.
For $\gamma<1$, on the other hand, 
the system described by Eq.~(\ref{ccjjcg:AFHM}) is equivalent to 
the quantum XY chain, where the Mermin-Wagner theorem prohibits
genuine long-range order.  In this case, the system
can be mapped to the repulsive Luttinger model, and 
the transverse component of the magnetization as well as
the $z$-component of the staggered magnetization exhibits
quasi-long-range order.
Namely, both $\avgl{(-1)^xS^z(x)S^z(0)}\avgr$ and
$\avgl{S^+(x)S^-(0)}\avgr$ decay algebraically with the distance $x$,
and the system in the charge picture displays
{\em both} the diagonal and the off-diagonal quasi-long-range order.  This
state may be regarded as the counterpart of the supersolid, possessing both
the diagonal and off-diagonal (true) long-range order and proposed recently
in 2D Josephson-junction arrays~\cite{Bruder93}.

The properties of the repulsive Luttinger liquid phase, with
both the diagonal and off-diagonal quasi-long-range order, has been
discussed in Ref.~\onlinecite{Glazma97} for a {\em single} Josephson-junction
chain.  It has been suggested that the system is extremely sensitive to
impurities~\cite{Kanexx92b} and may make another insulator, 
different in nature from the CDW insulator.
In the coupled chains, the repulsive Luttinger liquid phase has another
remarkable feature of the current mirror.
According to the basic transport mechanism due to
the cotunneling process of particles in the two chains,
shown in Fig.~\ref{ccjjcg:fig2},
the current fed through one chain is accompanied by the
secondary current in the other chain, with the same magnitude but in the
opposite direction.  
Similar current mirror effects have also been pointed out for $n_g=0$,
where the mechanism is rather different and
via the particle-hole pair transport (see below and
Ref.~\onlinecite{mschoi97p5}).


We now turn to the region near the particle-hole symmetry
line.  In the reduced Hilbert space $\varE_s$ with the condition
$n_+(x)=0$ satisfied, $n_-(x)/2$ can be
regarded as the number of particle-hole pairs located at $x$.
The role of such particle-hole pairs can be analyzed by means of the
imaginary-time path-integral representation of the partition function and
its dual transformation~\cite{mschoi97p5}. 
The Euclidean action, in the current-loop representation~\cite{Otterl94},
then reads
\onecolm
\begin{equation}
\varS
= \frac{1}{4K}\sum_{\ell\ell',xx',\tau}[n_\ell(x,\tau)-n_g]
    \left[2C_I\BfC^{-1}_{\ell\ell'}(x,x')\right]
    [n_\ell(x',\tau)-n_g]
  + \frac{1}{4K}\sum_{\ell,x,\tau}[J_\ell(x,\tau)]^2
  ,
\label{ccjjcg:a5}
\end{equation}
\twocolm\noindent
where the (imaginary) time has been rescaled in units of the inverse
Josephson plasma frequency $\omega_p^{-1}\equiv\hbar/\sqrt{4E_IE_J}$,
the dimensionless coupling constant defined to be
$K\equiv\sqrt{E_J/16E_I}$.
Here $(n_\ell, J_\ell)$ may be viewed as the current in
($1{+}1$)-dimensions, satisfying the continuity equation
\begin{equation}
\nabla_\tau{}n_\ell(x,\tau)+\nabla_xJ_\ell(x,\tau) = 0.
\label{ccjjcg:continuity}
\end{equation}
With the capacitance matrix Eq.~(\ref{ccjjcg:CC}), it is
convenient to decompose the action in Eq.~(\ref{ccjjcg:a5}) into the sum
$S=S_++S_-$:
\begin{eqnarray}
\varS_+
& \simeq & 
  \frac{E_0}{4KE_I}
    \sum_{x,\tau}[n_+(x,\tau)-2n_g]^2
  + \frac{1}{8K}\sum_{x,\tau}[J_+(x,\tau)]^2
  \nonumber \\
\varS_-
& \simeq & \frac{1}{8K}\sum_{x,\tau}[n_-(x,\tau)]^2
  + \frac{1}{8K}\sum_{x,\tau}[J_-(x,\tau)]^2
  \label{ccjjcg:varS-}
\end{eqnarray}
where $J_\pm(x,\tau)\equiv{}J_1(x,\tau)\pm J_2(x,\tau)$.

The factor $E_0/E_I$ in the component $\varS_+$, which is enormous in the
parameter regime of interest, again implies
the condition $n_+(x)=0$ already mentioned.
Further, the continuity equation in Eq.~(\ref{ccjjcg:continuity}) 
requires $J_+(x)$ to be a constant on the average, 
which should obviously be zero.  Consequently, near
the transition point, the system is effectively described by the action
$\varS_-$ in Eq.~(\ref{ccjjcg:varS-}), which is equivalent to the 2D XY
model, and exhibits a
BKT transition at
$K=K_{BKT}\approx 2/\pi$~\cite{Berezi71}. 
Here the transition, which is between the Mott insulating phase and 
the superconducting phase, 
is driven exclusively by the particle-hole pairs 
represented by the variable $n_-(x)$, 
whereas $n_+$ and $J_+$ merely renormalize the action $\varS_-$ and
shift slightly the transition point~\cite{mschoi97p5}.
This shift of the transition point depends on the external charge $n_g$ 
and may be estimated in the following way:
The transition to the superconducting phase
occurs when the Josephson-coupling energy $E_J$
also overcomes the Coulomb blockade associated with a particle-hole pair.
Since the Coulomb blockade increases with $n_g$, 
approximately given by $8E_0n_g^2{+}4E_I$, the critical value of
$E_J$ is concluded to grow from the symmetry-line value $16K_{BKT}^2 E_I$ 
as $n_g$ is increased.
It is also stressed that the BKT-type transition
survives the gate voltage as long as the induced charge is sufficiently
small ($|n_g|\ll1/4$); this is in sharp contrast to the single-chain
case, where breaking the particle-hole symmetry by nonzero $n_g$ 
immediately alters
the universality class of the transition~\cite{Bruder93}.


In the parameter regimes other than those considered above,
the behavior of the system may be inferred by the following argument:
First, it is obvious that for $E_J\gg{}E_0$, 
the system should be a superconductor 
with each chain superconducting separately.  
Note that this superconducting phase, denoted by S, comes from the {\em
particle} (Cooper pair) transport as usual, thus different in character
from the superconducting phase in the region $E_J\ll{}E_0$.
In the latter, denoted by S$'$, only the coupled chains as a whole
is superconducting, with superconductivity arising from 
the {\em particle-hole pair} transport.
Far away from both the particle-hole symmetry line and the maximal
frustration lines, the Hamiltonian may be
projected onto the subspace where $n_1(x)=0,1$ and $n_2(x)=0,1$ for all
$x$, and single-particle processes dominate the transport in the system.

The observations so far are summarized by
the phase diagram displayed schematically in Fig.~\ref{ccjjcg:fig3}.
The phase transitions of our main concern 
are represented by the thick solid lines, separating
the CDW from the repulsive Luttinger liquid (LL) 
and the Mott insulator (MI) from the superconductor (S$'$);
the somewhat speculative boundaries discussed above are
depicted by dashed lines.
Here it is not clear within our approach 
whether the boundary between the repulsive Luttinger
liquid region and the superconducting region in the phase diagram 
describes a phase transition or merely a crossover.  
Furthermore, even in the single-chain case,
the properties of the repulsive Luttinger liquid phase is controversial,
and the possibility of an intermediate {\em normal} phase has recently
been raised as well~\cite{Kuhner97}.


Coupled chain systems can presumably be realized in experiment by
current techniques, which have already made it possible to fabricate
submicron metallic junction arrays with large inter-array
capacitances~\cite{Matter97} as well as large arrays of
ultra-small Josephson junctions~\cite{Graber92}.
We also point out that quasiparticles have been safely disregarded
in obtaining the equilibrium properties at zero temperature.


This work was supported in part by the Ministry of Science and Technology
through the CRI Program, from the Ministry of Education through the BSRI
Program, and from the KOSEF through the
SRC Program.


\begin{thebibliography}{10}
\bibitem{ccjjcg:e2}
The minus sign in front of the second term in Eq.~(\ref{ccjjcg:AFHM}) is
not crucial here on a biparticle lattice (see Ref.~\onlinecite{Auerba94}).
\end{thebibliography}

\narrowtext 

\begin{figure}
\begin{center}
\epsfig{file=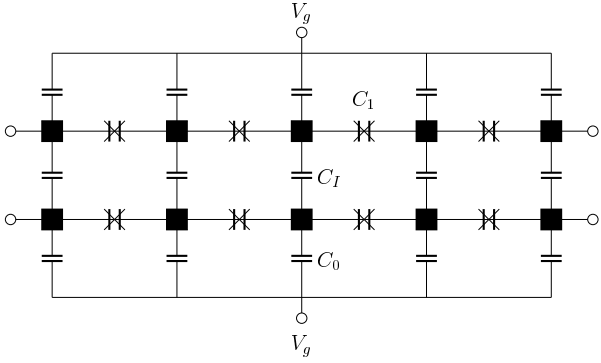,clip=,width=0.95\columnwidth}
\end{center}
\caption{Schematic diagram of the system.} 
\label{ccjjcg:fig1}
\end{figure}

\begin{figure}
\begin{center}
\epsfig{file=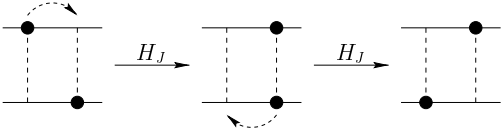,clip=,width=0.8\columnwidth}
\end{center}
\caption{A typical second-order process via intermediate virtual states
with energies of the order of $E_0$ near the maximal-frustration line.}
\label{ccjjcg:fig2}
\end{figure}

\begin{figure}
\begin{center}
\epsfig{file=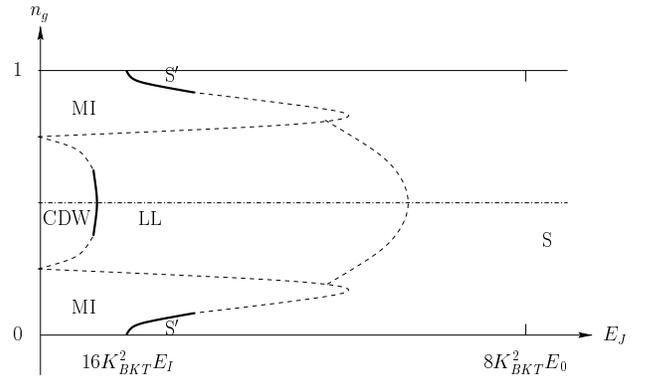,clip=,width=0.95\columnwidth}
\end{center}
\caption{Schematic phase diagram of the coupled Josephson-junction chains.}
\label{ccjjcg:fig3}
\end{figure}

\end{multicols}

\end{document}